\documentclass[12pt]{article}
\usepackage{epsfig}
\newcommand{\figwidth}{10.8cm}
\newcommand{\gapprox}{{\raisebox{-0.5ex}{${\scriptstyle>}$} \atop 
\raisebox{0.5ex}{${\scriptstyle\sim}$}}}
\newcommand{\lapprox}{{\raisebox{-0.5ex}{${\scriptstyle<}$} \atop 
\raisebox{0.5ex}{${\scriptstyle\sim}$}}}

\begin{document}

\title{Neutron stars and quark phases%
\addtocounter{footnote}{1}\footnote{Supported by DFG and GSI Darmstadt.}
 \\ in the NJL model%
\addtocounter{footnote}{-2}\footnote{This paper forms part of the 
dissertation of K.~Schertler.}}

\author{Klaus Schertler%
\addtocounter{footnote}{1}\footnote{E-mail: klaus.schertler@theo.physik.uni-giessen.de} 
~and Stefan Leupold \\
{\small \it Institut f\"ur Theoretische Physik, Universit\"at Giessen }\\
{\small \it D-35392 Giessen, Germany } \\
\smallskip \\
J\"urgen Schaffner-Bielich  \\
{\small \it RIKEN BNL Research Center, Brookhaven National Laboratory, }\\
{\small \it Upton, New York 11973-5000, USA } }
\date{}
\maketitle
\begin{figure}[ht]
\centerline{\epsfig{file=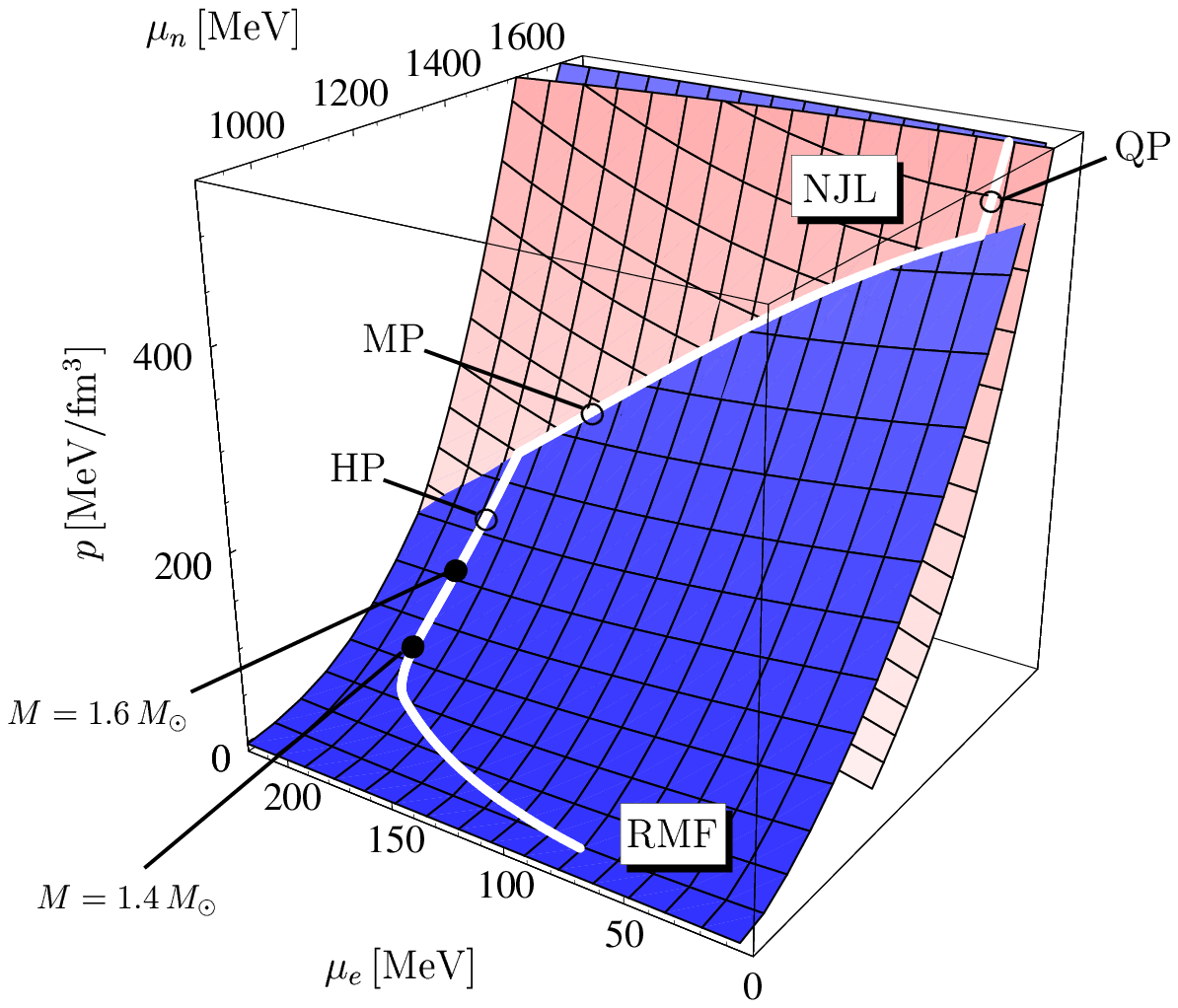,width=5cm}}
\end{figure}
\begin{abstract}
We study the possible existence of deconfined quark matter in the interior of 
neutron stars using the Nambu--Jona-Lasinio model to describe the quark phase. 
We find that typical neutron stars with masses around 1.4 solar masses 
do not possess
any deconfined quark matter in their center. This can be traced back to the 
property of the NJL model which suggests a large
constituent strange quark mass over a wide range of densities.
\bigskip
 
\end{abstract}

\section{Introduction}

At large temperatures or large densities hadronic matter is expected to undergo two phase 
transitions: one which deconfines quarks (and gluons) and one which restores chiral 
symmetry. Up to now it is an unsettled issue whether these two phase transitions are 
distinct or coincide. The more, it is even unclear whether there are real phase transitions
or only rapid crossover transitions. 
Such transitions have received much attention in heavy
ion physics as well as in the context of neutron stars which provide a unique environment to
study cold matter at supernuclear densities \cite{ShapiroBook,GlenBook}.
Even though a deconfinement phase transition seems intuitively evident at large 
enough densities, from a theoretical point of view a confirmation of the existence 
of a deconfined quark phase in neutron stars 
is so far limited by the uncertainties in modeling QCD at large densities. 
All the more it is important
to study and compare different available models to shed some light on 
similarities and differences
with respect to the behavior of matter at large densities as well as on the 
corresponding predictions of 
neutron star properties like e.g.~its mass and radius. In the future such 
experience may prove to be useful if either 
an improved understanding of matter under extreme conditions provides a more exclusive 
selection between 
the various models or new experimental results on neutron star properties are available 
to set more stringent 
constraints.

Usually the quark matter phase is modeled in the context of the MIT bag model 
\cite{GlenBook,MITbag,FarhiJaffe84}
as a Fermi gas of  $u$, $d$, and $s$ quarks. In this model the phenomenological bag 
constant $B_{\rm MIT}$ is
introduced to mimic QCD interactions to a certain degree. The investigation of such a 
phase was furthermore stimulated by the idea that a quark matter phase
composed of almost an equal amount of the three lightest quark flavors could be the 
ground state of nuclear matter
\cite{GlenBook,FarhiJaffe84,Bodm71,Witte84,MadsHaen91}. 
Indeed, for a wide range of model parameters such as the bag constant, 
bag models predict that the quark matter 
phase is absolutely stable i.e.~its energy per baryon at zero pressure
is lower than the one of $^{56}$Fe. If this is true, this has important
consequences in physics and astrophysics \cite{MadsHaen91} leading e.g.~to the 
possibility of so 
called ``strange stars'' \cite{GlenBook,MadsHaen91} which are neutron stars 
purely consisting of 
quark matter in weak equilibrium with electrons. Of course, to check the model dependence
of such findings it is important to perform the corresponding calculations also in models
different from the MIT bag model. 
In a recent work by Buballa and Oertel \cite{BO98} the equation of state (EOS) of 
quark matter was investigated
in the framework of the Nambu--Jona-Lasinio (NJL) model with three quark flavors. 
Applying this model it was found
that strange quark matter is not absolutely stable. This would rule out the existence of 
strange stars.
On the other hand, the possibility of quark phases in the interior of 
neutron stars is in principle 
not excluded by this result -- even though this possibility gets 
energetically less likely. 
Only a detailed phase transition calculation can answer the question which effect 
the findings in \cite{BO98}
have on the existence of quark phases inside neutron stars. 
This is what we are aiming at in the present work. 

In principle, for the description of a neutron star which consists of a quark phase in its
center and a surrounding hadronic phase (and, as we shall discuss below, 
a mixed phase in between) we need models for both phases. The most favorite case would
be to have one model which can reliably describe both phases. So far, there are no such models.
Therefore, we will use various versions of the relativistic mean field model to parametrize
the hadronic phase. For the quark phase we follow Buballa and Oertel \cite{BO98} in using the 
three-flavor version of the NJL model. The NJL model has proved to be very successful in the
description of the spontaneous breakdown of chiral symmetry exhibited by the true
(nonperturbative) QCD vacuum. It explains very well the spectrum of the low lying mesons which
is intimately connected with chiral symmetry as well as many other low energy phenomena of 
strong interaction \cite{VW91,Kl92,HK94}. 
At high enough temperature and/or density the NJL model 
predicts a transition to a state where chiral symmetry becomes restored. 
Despite that promising features which at first sight might suggest the NJL model as a good
candidate for modeling both the low and high density region of a neutron star this model
has one important shortcoming, namely it does not confine quarks. At low densities, however,
the bulk properties of strongly interacting matter are significantly influenced by the fact
that quarks are confined there. Therefore, we cannot expect that the NJL model gives 
reliable results for the EOS at low densities. Thus we will use the relativistic 
mean field model to describe the confined phase. At higher densities, however,
the quarks are expected to be deconfined. There we expect the NJL model to be applicable
since the lack of confinement inherent to this model is irrelevant in that regime. 
The interesting feature of the NJL model is that it reflects the chiral symmetry
of QCD. Clearly, it would be preferable 
to have a Lagrangian for the hadronic phase which also respects chiral symmetry
like e.g.~the one constructed in \cite{FTS95} for the two-flavor case and the SU(3) 
generalizations \cite{PZ98,Mu98}. 
Such Lagrangians, however, are more complicated to deal with.
First applications to neutron star matter seem to indicate that the modifications are 
rather small as compared to the relativistic mean field models used here \cite{priv}.
For simplicity, we therefore will restrict our
considerations to the much simpler extensions of the Walecka model 
which include hyperonic degrees of freedom (relativistic mean field
models). 

The paper is organized as follows: In Sec.~\ref{sec:HP} we discuss how the EOS 
for the hadronic
phase of a neutron star is calculated within several variants of relativistic mean field
models. We keep brief here since such models are frequently used and well documented
in the literature (cf.~e.g.~\cite{GlenBook}).
In Sec.~\ref{sec:QP} we apply the NJL model to the description of the 
possible quark phase of the neutron star. Here we present much more details as compared to 
Sec.~\ref{sec:HP} since to the best of our knowledge it is the first time that the
NJL model is applied to the description of the quark phase in a neutron star.
Sec.~\ref{sec:PT} is devoted to the construction
of the phase transition and to the application of the complete EOS to the
internal structure of the neutron star. Finally we summarize and discuss our results in 
Sec.~\ref{sec:sum}.

\section{Hadronic matter}  \label{sec:HP}
Neutron stars cover a wide range of densities. From the surface of the star 
which is composed
of iron with a density of $\epsilon \approx 8 \,$g/cm$^3$ the density can increase 
up to several times
normal nuclear matter density \mbox{($\epsilon_0 = 140\,$MeV/fm$^3$} 
\mbox{$\approx 2.5 \times 10^{14}\,$g/cm$^3$)} in the center of the star. Since 
there is no single
theory that covers this huge density range, we are forced to use different 
models to meet the
requirements of the various degrees of freedom opened up at different densities.
For subnuclear densities we apply the Baym-Pethick-Sutherland EOS \cite{BPS}. 
The degrees of freedom in this EOS are nuclei, electrons and neutrons. The 
background of neutrons appears above neutron drip density  
($\epsilon_{\rm drip}\approx$ \mbox{$4\times 10^{11}\,$g/cm$^3$}) when the most 
weakly bound neutrons 
start to drip out of the nuclei which themselves get more and more neutron rich 
with increasing density. 
For a detailed discussion of the Baym-Pethick-Sutherland EOS see also 
\cite{ShapiroBook}. We also refer to \cite{STOS98} where a relativistic mean field
model is extended to also describe this low density range. 

At densities of about normal nuclear density $\epsilon_0$ the nuclei begin to 
dissolve and 
merge together and nucleons become the relevant degrees of freedom in this phase. 
We want to 
describe this phase in the framework of the relativistic mean field (RMF) model 
which
is widely used for the description of dense nuclear matter 
\cite{RMF,Glen8287,SchaMish96}.
For an introduction to the RMF model see e.g.\,\cite{GlenBook}. 
We use three EOS's calculated by Schaffner and Mishustin in the extended RMF model 
\cite{SchaMish96} (denoted as TM1, TM2, GL85) and one by Gosh, Phatak and Sahu
\cite{Gosh95}. For the latter one we use GPS as an abbreviation. These models 
include hyperonic
degrees of freedom which typically appear at 
$\epsilon \approx 2 \! - \! 3$ $\epsilon_0$. 
Table \ref{NuclProps} 
shows the nuclear matter properties and the particle composition of the four EOS's. 
%
\begin{table}
\begin{center}
\begin{tabular}{ccccc}
\hline
Hadronic EOS & TM1 & TM2 & GL85 & GPS \\
\hline
reference & \cite{SchaMish96} & \cite{SchaMish96} & \cite{SchaMish96} & 
\cite{Gosh95} \\  
$\rho_0\,$[fm$^{-3}$] & $0.145$ & $0.132$ & $0.145$ & $0.150$ \\ 
$B/A\,$[MeV] & $-16.3$ & $-16.2$ & $-15.95$ & $-16.0$ \\
$K\,$[MeV] & $281$ & $344$ & $285$ & $300$ \\
$m_N^*/m_n$ & $0.634$ & $0.571$ & $0.770$ & $0.830$ \\
$a_{sym}\,$[MeV] & $36.9$ & $35.8$ & $36.8$ & $32.5$ \\
composition & a & a & a & b \\
\hline
\multicolumn{5}{l}{a) n, p, e$^-$, $\mu^-$, $\Lambda$, $\Sigma^-$, $\Sigma^0$, 
$\Sigma^+$,
$\Xi^-$, $\Xi^0$}\\
\multicolumn{5}{l}{b) n, p, e$^-$, $\mu^-$, $\Lambda$, $\Sigma^-$}\\
\hline
\end{tabular}
\end{center}
\caption{Nuclear matter properties of the hadronic EOS's. The saturation density and
the binding energy is denoted by $\rho_0$ and $B/A$, the incompressibility by $K$, 
the
effective mass by $m_N^*/m_n$ and the symmetry energy by $a_{sym}$. The particle 
compositions
are shown at the bottom of the table.}
\label{NuclProps}
\end{table}
%
The RMF EOS's are matched to the Baym-Pethick-Sutherland EOS at densities of 
$\epsilon \approx 10^{14}\,$g/cm$^3\approx \epsilon_0$. Even if the relevant 
degrees 
of freedom are specified (in the RMF case basically nucleons and hyperons) 
the high density 
range of the EOS is still not well 
understood. The use of different hadronic models should reflect this uncertainty 
to some degree.
In the following we denote the phase described by the Baym-Pethick-Sutherland EOS 
and by the RMF model as the {\em hadronic phase} (HP) of the neutron star.

\section{Quark phase}  \label{sec:QP}

To describe the deconfined {\it quark phase} (QP) we use the Nambu--Jona-Lasinio 
(NJL) model 
\cite{NJL} with three flavors \cite{RKH96} in Hartree (mean field) approximation 
(for reviews on the NJL model cf.~\cite{VW91,Kl92,HK94}). 
The Lagrangian is given by 
(cf.~\cite{BO98,RKH96})
\begin{eqnarray}
{\cal L} &=& \bar q \, ( i \! \! \not \! \partial - {\hat m}) \, q
            + G \sum_{k=0}^8 [\,(\bar q \lambda_k q)^2 + 
           (\bar q i\gamma_5\lambda_k q)^2\,] 
\nonumber \\  && \phantom{mmmmmm}           
- K \,[ \,{\rm det}_f (\bar q \, (1+\gamma_5) \, q) 
           + {\rm det}_f (\bar q \, (1-\gamma_5) \, q) \,]   
\label{eq:nambulag}
\end{eqnarray}
where $q$ denotes a quark field with three flavors, $u$, $d$, and $s$, and
three colors. ${\hat m} = {\rm diag}(m_{u},m_{d},m_{s})$ is a 
$3 \times 3$ matrix 
in flavor space. For simplicity we use the isospin symmetric case, 
$m_{u}=m_{d} \equiv m_{q}$. The $\lambda_k$ matrices act in flavor space.
For $k = 1, \dots , 8$ they are the generators of $SU(3)_f$ while 
$\lambda_0$ is proportional to the unit matrix in flavor space (see \cite{Kl92}
for details). The four-point interaction term $\sim G$ is symmetric in 
$SU_V(3) \times SU_A(3) \times U_V(1) \times U_A(1)$. 
In contrast, the determinant term $\sim K$
which for the case of three flavors generates a six-point interaction breaks
the $U_A(1)$ symmetry. If the mass terms are neglected the overall symmetry of 
the Lagrangian therefore is 
$SU_V(3) \times SU_A(3) \times U_V(1)$. In vacuum this symmetry is spontaneously
broken down to $SU_V(3) \times U_V(1)$ which implies the strict conservation of baryon and
flavor number. The full chiral symmetry -- which implies in addition the conservation of the 
axial flavor current -- becomes restored at sufficiently high
temperatures and/or densities. The finite mass terms introduce an additional
explicit breaking of the chiral symmetry. On account of the chiral symmetry breaking mechanism the quarks
get constituent quark masses which in vacuum are considerable larger than their current quark mass 
values. In media with very high quark densities constituent and current quark masses become approximately
the same (concerning the strange quarks this density regime lies far beyond the point where chiral
symmetry is restored).

The coupling constants $G$ and $K$ appearing in (\ref{eq:nambulag}) have
dimension energy$^{-2}$ and energy$^{-5}$, respectively.
To regularize divergent loop integrals we use for simplicity a sharp cut-off
$\Lambda$ in 3-momentum space. Thus we have at all five parameters, namely
the current quark masses $m_{q}$ and $m_{s}$, 
the coupling constants $G$ and $K$,
and the cut-off $\Lambda$. Following \cite{RKH96} we use $\Lambda = 602.3\,$MeV,
$G \Lambda^2 = 1.835$, $K \Lambda^5 = 12.36$, $m_{q} = 5.5 \,$MeV, and 
$m_{s} = 140.7\,$MeV. These parameters are chosen such that the empirical values 
for the pion decay constant and the meson masses of pion, kaon and $\eta'$ can
be reproduced. The mass of the $\eta$ meson is underestimated by about 6\%.

We treat the three-flavor NJL model in the Hartree approximation which amounts
to solve in a selfconsistent way the following gap equations for the dynamically 
generated constituent (effective) quark masses:
\begin{equation}
  \label{eq:gap}
m_i^* = m_{i} - 4 G \langle \bar q_i q_i \rangle 
+ 2 K \langle \bar q_j q_j \rangle \langle \bar q_k q_k \rangle
\end{equation}
with $(i,j,k)$ being any permutation of $(u,d,s)$. At zero temperature but 
finite quark chemical potentials the quark condensates
are given by
\begin{equation}
  \label{eq:cond}
\langle \bar q_i q_i \rangle = - 2 N_c 
\int\limits_{p_F^i < \vert \vec p \vert < \Lambda} \!\! {d^3 \! p \over (2 \pi)^3}
\, {m_i^* \over \sqrt{(m_i^*)^2 + \vec p\,^2}} 
= - {3 \over \pi^2} \int \limits_{p_F^i}^\Lambda \!\! dp \, p^2 \, 
{m_i^* \over \sqrt{(m_i^*)^2 + p^2}} 
\end{equation}
where we have taken the number of colors to be $N_c = 3$. 
$p_F^i$ denotes the Fermi momentum of the respective quark flavor $i$. It is
connected with the respective quark chemical potential $\mu_i$ via
\begin{equation}
  \label{eq:fermimom}
p_F^i = \sqrt{\mu_i^2 - (m_i^*)^2 } \, \Theta(\mu_i - m_i^*)   \,.
\end{equation}
The corresponding quark particle number density is given by
\begin{equation}
  \label{eq:numbdens}
\rho_i = 2 N_c 
\int\limits_{\vert \vec p \vert < p_F^i} \!\! {d^3 \! p \over (2 \pi)^3}
= {(p_F^i)^3 \over  \pi^2 }  \,.
\end{equation}
For later use we also introduce the baryon particle number density
\begin{equation}
  \label{eq:barypartdens}
\rho \equiv {1 \over 3} (\rho_u + \rho_d + \rho_s )   \,. 
\end{equation}
The Eqs.~(\ref{eq:gap}), (\ref{eq:cond}) serve to generate constituent quark masses
which decrease with increasing densities from their vacuum values of 
$m^*_{q,\rm vac} = 367.7\,$MeV and $m^*_{s,\rm vac} = 549.5\,$MeV, respectively. 

Before calculating the EOS we would like to comment briefly on the Hartree 
approximation to the NJL model which we use throughout this work. 
This treatment is identical to a leading order calculation in the inverse
number of colors $1/N_c$ \cite{Kl92}. 
In principle, one can go beyond this approximation by taking into account
$1/N_c$ corrections in a systematic way. This amounts in the inclusion of quark-antiquark
states (mesons) as RPA modes in the thermodynamical calculations \cite{HKZV94,ZHK94}. 
Several things then change: First of all, these mesons might contribute to the EOS. 
We are not aware of a thorough discussion of such an EOS for three flavors with finite 
current quark masses. The two-flavor case is discussed in \cite{ZHK94}. 
Qualitatively the masses of the meson states rise above the chiral transition point.
Therefore, they should become energetically disfavored and thus less important. 
An additional technical complication arises due to the fact that the relation between the 
Fermi energy and the chemical potential becomes nontrivial. Instead of
(\ref{eq:fermimom}) one has to solve an additional gap equation for each flavor species.
These gap equations are coupled to the gap equations for the constituent quark masses
given in (\ref{eq:gap}). We refer to \cite{Kl92} for details. 
For simplicity we will restrict ourselves in the following to the Hartree 
approximation and comment on the possible limitations of that approach in the last
section. 

\begin{figure}
\centerline{\epsfig{file=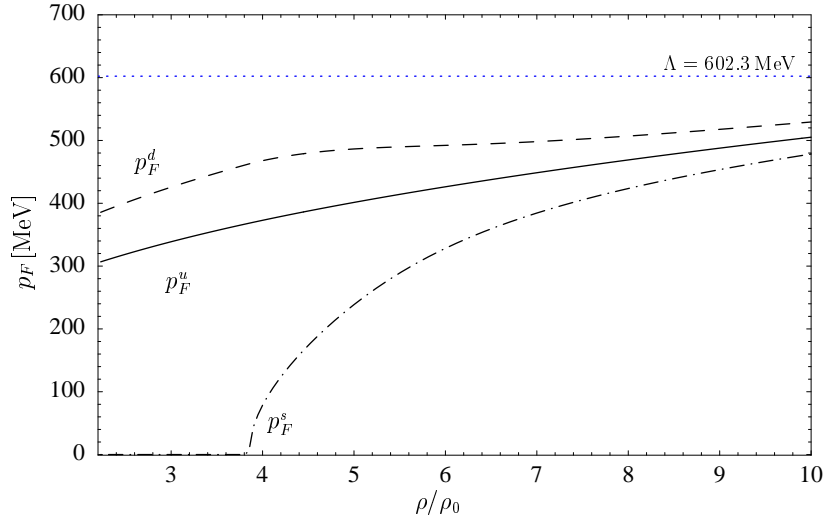,width=\figwidth}}
\caption{Fermi momenta of the quarks
as a function of baryon particle number density for a charge neutral system of 
quarks and electrons in weak equilibrium. 
$\Lambda$ denotes the cut-off introduced to
regularize the NJL model. $\rho_0 = 0.17\,$fm$^{-3}$ denotes nuclear saturation
density.}
\label{fig:pfRho}
\end{figure}

Coming back to the EOS we also need the energy density and the pressure of the 
quark system. 
In the Hartree approximation the energy density turns out to be
\cite{BO98}
\begin{equation}
  \label{eq:endensq}
\epsilon_{\rm NJL} = \sum\limits_{i = u,d,s} {3 \over \pi^2} 
\int\limits_0^{p_F^i} \!\! dp \, p^2 \sqrt{(m_i^*)^2 + p^2}  \, + B_{\rm eff}  
\end{equation}
while pressure and energy density are related via
\begin{equation}
  \label{eq:pressq}
p_{\rm NJL}  + \epsilon_{\rm NJL} = \sum\limits_{i = u,d,s} \rho_i \, \mu_i  
\end{equation}
where the effective bag pressure $B_{\rm eff}$ is given by
\begin{equation}
  \label{eq:defeffbag}
B_{\rm eff} = B_0 - B 
\end{equation}
with
\begin{eqnarray}
B & = & \sum\limits_{i = u,d,s}  
\left[
{3 \over \pi^2} \int\limits_0^{\Lambda} \!\! dp \, p^2
\left( \sqrt{(m_i^*)^2 + p^2} - \sqrt{(m_i)^2 + p^2}  \right)
- 2 G \langle \bar q_i q_i \rangle^2
\right] 
\nonumber  \\  && {} 
+ 4 K \langle \bar u u \rangle \langle \bar d d \rangle \langle \bar s s \rangle    
  \label{eq:defbag}
\end{eqnarray}
and
\begin{equation}
  \label{eq:defb0}
B_0 = \left. B \right\vert_{\rho_u = \rho_d = \rho_s = 0} = (217.6 \,\mbox{MeV})^{4} \,\,.
\end{equation}
Note that $B$ depends implicitly on the quark densities via the (density dependent) 
constituent quark masses. The appearance of the density independent constant $B_0$
ensures that energy density and pressure vanish in vacuum. We note here that
this requirement fixes the density independent part of $B_{\rm eff}$ 
which influences the EOS via (\ref{eq:endensq}), (\ref{eq:pressq}) and therefore the 
possible phase transition to quark matter. We will come back to this
point in the last section. In what follows
we shall frequently compare the results of the three-flavor NJL model with the
simpler MIT bag model \cite{MITbag,FarhiJaffe84}. For that purpose it is important
to realize that the NJL model predicts a (density dependent) bag pressure
$B_{\rm eff}$ while in the MIT bag model the bag constant $B_{\rm MIT}$ 
is a density independent
free parameter. There usually also the quark masses $m_i^{\rm MIT}$ 
are treated as density 
independent quantities. (An exception is the model discussed in 
\cite{Sche97,Sche98} which uses density dependent effective quark masses
caused by quark interactions in the high density regime.)
In the bag model energy density and pressure of the quark
system are given by
\begin{equation}
  \label{eq:endensMIT}
\epsilon_{\rm MIT} = \sum\limits_{i = u,d,s} {3 \over \pi^2} 
\int\limits_0^{p_F^i} \!\! dp \, p^2 \sqrt{(m_i^{\rm MIT})^2 + p^2}  \, + 
B_{\rm MIT}
\end{equation}
and
\begin{equation}
  \label{eq:pressMIT}
p_{\rm MIT} + \epsilon_{\rm MIT} = \sum\limits_{i = u,d,s} \rho_i \, \mu_i    \,.
\end{equation}
Suppose now that the densities are so high that in the three-flavor NJL model the 
effective quark masses have dropped down to the current quark masses. In this case, energy 
density and pressure take the form of the respective expressions in the MIT bag 
model with $m_i^{\rm MIT} = m_i$ and $B_{\rm MIT} = B_0$. However, a word of
caution is in order here. For {\it very high} quark particle number densities the
corresponding Fermi momenta become larger than the momentum cut-off $\Lambda$
introduced to regularize the NJL model. In this case the results of the NJL
model become unreliable. E.g.~the upper limit of the momentum integration
in (\ref{eq:numbdens}) would be no longer given by the Fermi momentum but by the
cut-off $\Lambda$ which would be clearly an unphysical behavior of the model. 
Thus for all practical purposes one should always
ensure that in the region of interest the Fermi momenta are smaller than the 
momentum cut-off $\Lambda$. Fig.~\ref{fig:pfRho} shows the Fermi momenta of the
quarks as a function of the baryon particle number density (for a charge neutral
system of quarks and electrons in weak equilibrium; cf.~next paragraph for 
details). 
Obviously all Fermi momenta stay below the cut-off $\Lambda$ for the region
of interest. We will come back to that point at the end of this section.

The QP which might be found in the center of a neutron star consists of 
$u$, $d$, and $s$ quarks and electrons in weak
equilibrium, i.e.~the weak reactions
\begin{eqnarray}
d & \longrightarrow & u+e^- +\bar{\nu}_{e^-} \,, \nonumber \\
s & \longrightarrow & u+e^- +\bar{\nu}_{e^-} \,, \\
s + u & \longleftrightarrow & d + u  \nonumber
\end{eqnarray}
imply relations between the four chemical potentials 
$\mu_u$, $\mu_d$, $\mu_s$, $\mu_e$ which read
\begin{equation}
\mu_s  = \mu_d =  \mu_u+\mu_e  \,.
\label{eq:weakeq}
\end{equation}
Since the neutrinos can diffuse out of the star their chemical potentials are 
taken to be zero.
The number of chemical potentials necessary for the description of the QP in weak
equilibrium is therefore reduced to {\it two} independent ones. For convenience 
we choose the pair ($\mu_n$, $\mu_e$) with the neutron chemical potential
\begin{equation}
  \label{eq:defneutronpd}
\mu_n \equiv \mu_u + 2 \mu_d  \,. 
\end{equation}
In a pure QP (in contrast to quark matter in a mixed phase which we will 
discuss later) we can 
require the QP to be charge neutral. This gives us an additional constraint on the
chemical potentials via the following relation for the particle number densities:
\begin{equation}
\frac{2}{3} \rho_u -\frac{1}{3} \rho_d -\frac{1}{3} \rho_s 
-\rho_e = 0  
\label{eq:chargeneu}
\end{equation}
where $\rho_e$ denotes the electron particle number density. Neglecting the
electron mass it is given by
\begin{equation}
  \label{eq:elecpartdens}
\rho_e = { \mu_e^3  \over 3 \pi^2}  \,.
\end{equation}
Utilizing the relations (\ref{eq:weakeq}) and (\ref{eq:chargeneu})
the EOS can now be parametrized by only {\it one} chemical potential, say $\mu_n$. 
At this point it should
be noted that the arguments given here for the QP also holds for the HP. There 
one also ends up
with {\it two} independent chemical potentials (e.g.~$\mu_n$ and $\mu_e$)
if one only requires weak equilibrium
between the constituents of the HP and
with {\it one} chemical potential (e.g.~$\mu_n$) if one additionally 
requires charge neutrality. As we will 
discuss later, the number of independent chemical potentials plays a crucial role 
in the formulation
of the Gibbs condition for chemical and mechanical equilibrium between the HP and 
the QP. 

In the pure QP total energy density and pressure are given by the respective
sums for the quark and the electron system, i.e.
\begin{equation}
  \label{eq:enerQP}
\epsilon = \epsilon_{\rm NJL} + {\mu_e^4 \over 4 \pi^2}
\end{equation}
and
\begin{equation}
  \label{eq:pressQP}
p = p_{\rm NJL} + {\mu_e^4 \over 12 \pi^2}
\end{equation}
where the system of electrons is treated as a massless ideal gas. One obtains the analogous
expressions for the MIT bag model if $\epsilon_{\rm NJL}$ and $p_{\rm NJL}$ are
replaced by the respective MIT expressions (\ref{eq:endensMIT}) and 
(\ref{eq:pressMIT}).

\begin{figure}
\centerline{\epsfig{file=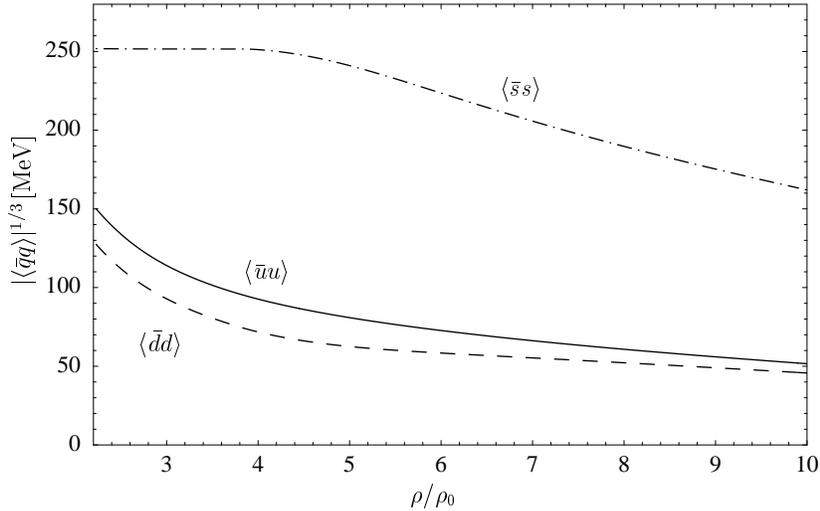,width=\figwidth}}
\caption{Quark condensates 
as a function of baryon particle number density for a charge neutral system of 
quarks and electrons in weak equilibrium. Note that all condensates have negative 
values (cf.~(\protect{\ref{eq:cond}})).}
\label{fig:qqRho}
\end{figure}

\begin{figure}
\centerline{\epsfig{file=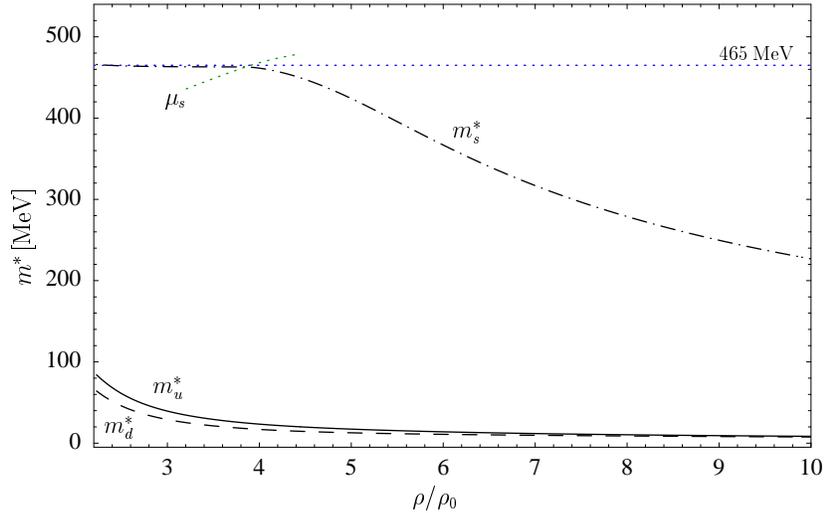,width=\figwidth}}
\caption{Effective quark masses 
as a function of baryon particle number density for a charge neutral system of 
quarks and electrons in weak equilibrium. In addition the strange quark
chemical potential $\mu_s$ is plotted in the region where it meets the constituent
strange quark mass. This marks the point where strange quarks appear in the system
(see also Fig.~\protect\ref{fig:pfRho}).}
\label{fig:mRho}
\end{figure}

\begin{figure}
\centerline{\epsfig{file=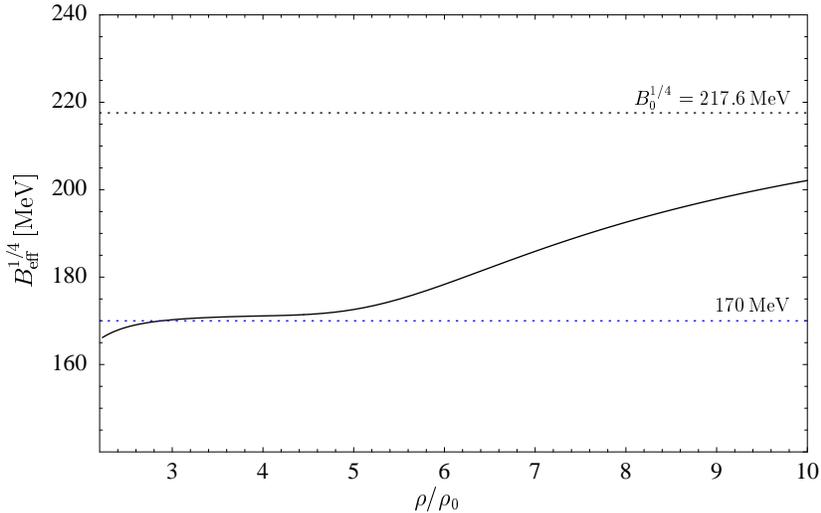,width=\figwidth}}
\caption{The bag pressure (to the power of 1/4) as function of the baryon particle 
number density for the same conditions as described in 
Fig.~\protect\ref{fig:qqRho} and \protect\ref{fig:mRho}.}
\label{fig:BRho}
\end{figure}

\begin{figure}[ht]
\centerline{\epsfig{file=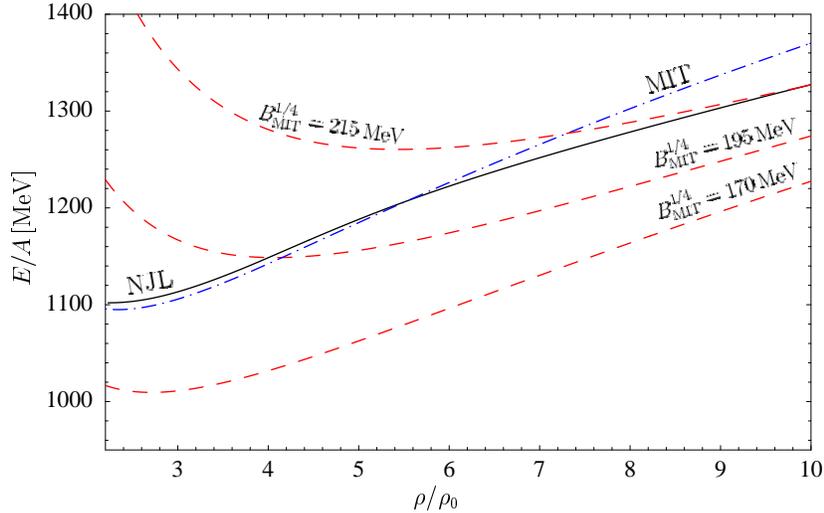,width=\figwidth}}
\caption{Energy per baryon versus baryon particle 
number density for the NJL model and the MIT bag model
for various values of the bag pressure and the strange quark mass. 
The curve labeled with ``MIT'' uses
a bag pressure of $B_{\rm MIT}^{1/4} = 170 \,$MeV and a strange quark mass
of $m_s^{\rm MIT} = 465 \,$MeV. All other bag model curves use the current strange 
quark mass of $140.7\,$MeV. See main text for details.}
\label{fig:EARho}
\end{figure}

\begin{figure}
\centerline{\epsfig{file=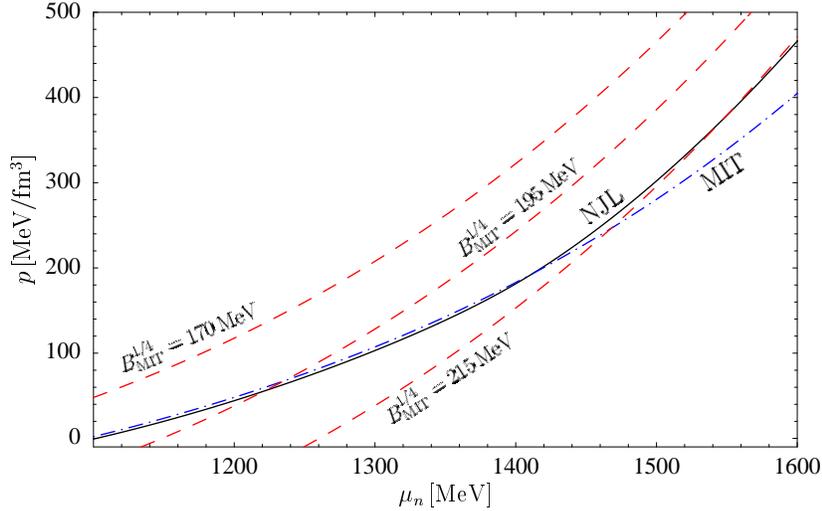,width=\figwidth}}
\caption{Pressure versus neutron chemical potential for the NJL model and the 
MIT bag model
for various values of the bag pressure and the strange quark mass. 
See Fig.~\protect\ref{fig:EARho} and main text for details.}
\label{fig:pMub}
\end{figure}

\begin{figure}
\centerline{\epsfig{file=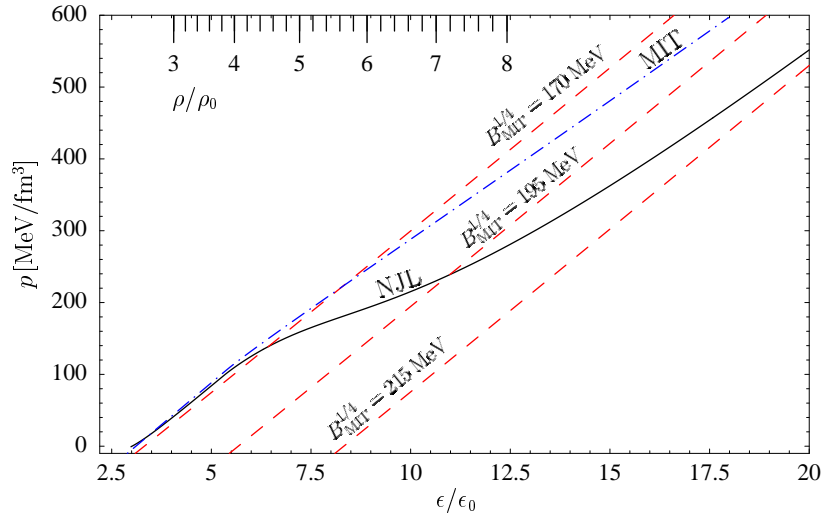,width=\figwidth}}
\caption{Pressure versus energy density for the NJL model and the MIT bag model
for various values of the bag pressure and the strange quark mass. 
The upper scale shows the corresponding particle number
densities. $\epsilon_0 = 140\,$MeV/fm$^3$. See Fig.~\protect\ref{fig:EARho} and 
main text for details.}
\label{fig:pEps}
\end{figure}

Demanding weak chemical equilibrium (\ref{eq:weakeq}) 
and charge neutrality (\ref{eq:chargeneu}) as discussed above 
all thermodynamic quantities as well as quark condensates, effective quark masses
etc.~can be calculated as a function of one chemical potential $\mu_n$. The 
curves in Fig.~\ref{fig:pfRho} as well as in Figs.~\ref{fig:qqRho}-\ref{fig:pEps} 
which we shall discuss in the
following are obtained  by varying $\mu_n$
while obeying simultaneously the constraints (\ref{eq:weakeq}) and 
(\ref{eq:chargeneu}). 

Figs.~\ref{fig:qqRho} and \ref{fig:mRho} show the quark condensates and the 
effective quark masses, respectively, as a function of the baryon
particle number density. 
Note that we start already at a density as high as
two times nuclear saturation density $\rho_0 = 0.17 \,$fm$^{-3}$ since we want to 
describe only the high density regime of the neutron star with quark degrees of
freedom while for low densities we use the hadronic EOS described in the previous
section. Concerning the low density regime of the three-flavor NJL model we refer 
to \cite{BO98} for details. 
There it was shown that the energy per baryon of a charge neutral 
system of quarks and electrons in weak equilibrium (described by the NJL model and 
a free electron gas) shows a minimum somewhat above two times $\rho_0$. This implies
that in the density region below this minimum the pressure is negative. We are
not interested in the (low density) part of the EOS with negative pressure since 
it cannot be realized in a neutron star. In the region of interest 
Figs.~\ref{fig:qqRho} and \ref{fig:mRho} show that the strange quark condensate 
and the effective 
strange quark mass stay constant until the strange chemical potential $\mu_s$ 
overwhelms the strange quark mass. Only then according to (\ref{eq:fermimom}) 
the strange quark particle number density $\rho_s$
and the corresponding Fermi momentum $p_F^s$ (cf.~Fig.~\ref{fig:pfRho})
become different from zero causing
a decrease of $\vert\langle \bar s s \rangle\vert$ and $m_s^*$. Note that all condensates 
have negative values (cf.~(\protect{\ref{eq:cond}})).
One might wonder why the 
dropping of the condensates of the light up and down quarks does 
not decrease the strange quark mass (and condensate) due to the last coupling term 
in (\ref{eq:gap}). Indeed, strange quark mass and condensate have 
dropped in the low density region (not shown here) from their vacuum values down to
the plateaus shown in Figs.~\ref{fig:qqRho}, \ref{fig:mRho} due to their coupling
to the up and down quark condensates. In the plateau region, however,
these condensates have already decreased so much that their
influence on the strange quark mass is diminished. We refer to \cite{BO98} for details. 
As we shall see below, the large plateau value of the strange quark mass will have
considerable influence on the phase structure in the interior of neutron stars. 

Fig.~\ref{fig:BRho} shows the bag pressure $B_{\rm eff}$ as a function of the baryon particle 
number density. 
After staying more or less constant up to roughly 5 times
nuclear saturation density it starts to increase towards $B_0$ which, however, it 
will reach only very slowly. Again the rising of $B_{\rm eff}$ can be traced back to the
strange quarks which come into play at high densities. 

Thermodynamic quantities are shown in Figs.~\ref{fig:EARho}-\ref{fig:pEps}.
For comparison various curves calculated within the MIT bag model are added.
The curves labeled with a specific value of the bag pressure are obtained from
(\ref{eq:endensMIT},\ref{eq:pressMIT}) in weak equilibrium
where the respective value of $B_{\rm MIT}$ and
the current quark masses $m_q = 5.5\,$MeV and $m_s = 140.7\,$MeV are used. 
In contrast to that for the curve 
labeled with ``MIT'' we have used the plateau values of the bag pressure 
$B_{\rm MIT}^{1/4} = 170 \,$MeV 
(cf.~Fig.~\ref{fig:BRho}) and of the strange quark mass $m_s^{\rm MIT} = 465 \,$MeV
(cf.~Fig.~\ref{fig:mRho}). For up and down quarks we have used the current 
quark mass values also here.
Fig.~\ref{fig:EARho} shows the energy per baryon as a function
of the baryon particle number density. 
We find that the results of the NJL
model calculation cannot be reproduced by a bag model using the current strange quark mass
-- no matter which bag pressure is chosen. As already discussed above, the reason simply is 
that in the NJL model up to four times
$\rho_0$ there are no strange quarks in a system which is in weak equilibrium 
(cf.~Fig.~\ref{fig:pfRho}). On the other hand, in bag models using the much lower current 
strange quark mass one finds a reasonable amount of strange quarks already at vanishing
pressure which typically corresponds to $2 \! - \! 3$ times
$\rho_0$ \cite{Sche97}. In contrast to that, a bag model with the plateau values for bag 
pressure and
strange quark mass (denoted as ``MIT'' in the figures) yields a very good approximation
to the NJL result for the energy per baryon up to $6 \! - \! 7$ times $\rho_0$. For higher
particle number densities the NJL result bends over and can be better described by
bag models using the current strange quark mass and higher bag pressures (roughly $B_0$).
All these findings also apply to the interpretation of Fig.~\ref{fig:pMub}
which shows the total pressure of the system versus the baryon chemical potential. 
Comparing the two curves with the same bag constant labeled with 
``$B_{\rm MIT}^{1/4} = 170\,$MeV'' and with ``MIT'', respectively, one observes
that the latter one has a significantly lower pressure. This is due to the use of
the much larger effective strange quark mass of $m_s^{\rm MIT} = 465 \,$MeV in the
latter case as compared to the current strange quark mass of $140.7\,$MeV used in
the former. The $p$ versus $\mu_n$ relation is an important ingredient for the 
construction of the phase transition from
hadronic to quark matter inside a neutron star. We note already here, however,
that we need in addition the thermodynamical relations also for a quark-electron system 
away from the charge neutral configuration to describe correctly the phase transition
(see below).

The outlined picture concerning the comparison of NJL and bag models is somewhat
modified when looking at Fig.~\ref{fig:pEps} which shows the total pressure as a function
of the energy density. 
EOS's in the form $p(\epsilon)$ enter the Tolman-Oppenheimer-Volkoff \cite{OppeVolk39} 
equation which in turn determines the mass-radius relation of neutron stars. 
We see that in the lower part of the plotted
energy density range the EOS in Fig.~\ref{fig:pEps} is 
reasonably well described by MIT bag models with the plateau value 
$B_{\rm MIT}^{1/4} = 170\,$MeV no matter which quark masses are chosen (current or
effective quark masses). The reason is that the $p(\epsilon)$ relation is not very
sensitive to the quark masses. This has already been observed in a somewhat different
context in \cite{Sche97}. Going to higher densities
the strange quarks enter the game and the EOS in Fig.~\ref{fig:pEps} obtained
from the NJL model starts to deviate from the EOS of the MIT bag models with 
the plateau value $B_{\rm MIT}^{1/4} = 170\,$MeV. 
For very high densities the pressure determined from the 
NJL model becomes comparable to the one calculated in the bag model with a high bag constant
(roughly $B_0$). It is interesting to note that the deviation between the NJL curve and
the ``MIT'' curve starts to increase in Fig.~\ref{fig:pEps} much earlier than in 
Figs.~\ref{fig:EARho} and \ref{fig:pMub}. 
This shows that the pressure versus energy density relation is much more sensitive to the
detailed modeling than the relations shown in Figs.~\ref{fig:EARho} and \ref{fig:pMub}. 

Before constructing the phase transition inside the neutron star 
let us briefly discuss the limitations of the NJL model in the form as we have treated 
it here. As a typical low energy theory the NJL model is not renormalizable. This is not an 
obstacle since such theories by construction should be only applied to low energy
problems. In practice the results depend on the chosen cut-off or, to turn the argument
around, the NJL model is only properly defined once a cut-off has been chosen. 
This cut-off serves as a limit for the range of applicability of the model. 
Here we
have used one cut-off $\Lambda$ for the three-momenta of all quark species. 
Concerning the discussion of other cut-off schemes and their interrelations we refer to 
\cite{Kl92,HK94}. When the density in the quark phase gets higher the Fermi momenta
of the quarks rise due to the Pauli principle. Eventually they might overwhelm the cut-off
of the NJL model. At least beyond that point the model is no longer applicable. We have made
sure in our calculations that this point is never reached (cf.~Fig.~\ref{fig:pfRho}).
In addition, at very high densities one presumably enters a regime which
might be better described by (resummed) perturbation theory. While the nonperturbative
features of the NJL model vanish with rising density, medium effects as mediated e.g.~by
one-gluon exchange grow with the density \cite{Sche97,Sche98}. 
To summarize, concerning the calculation of the EOS it turns out that the NJL model should
neither be used at low densities where confinement properties are important nor at
very high densities where the NJL model as a low energy theory leaves its range of
applicability. 
However, the NJL model might yield reasonable results in a window of the
density range where confinement is no longer crucial but chiral symmetry as a symmetry
of full QCD remains to be important.

\section{Phase transition and neutron stars} \label{sec:PT}
%
In the previous sections we have discussed the underlying EOS's thought to reflect 
the properties of confined hadronic matter (HP) and deconfined quark matter (QP) in
its particular regime of applicability. Applying these EOS's we want to calculate in this section
the phase transition from the HP to the QP to see which phase is the favored one at which densities.
(The existence of a QP inside the neutron star of course requires
the phase transition density to be smaller than the central density of the star.)

It is worth to point out which phase structure is in principle possible if a hadronic
model and the NJL model are connected at a certain density value $\rho_{\rm deconf}$ 
(which is dynamically determined in the present work by a Gibbs construction as we shall 
discuss below). At density $\rho_{\rm deconf}$ we assume a first order phase 
transition from confined hadronic to deconfined quark matter.
Even without a matching to a hadronic model
the NJL model already exhibits a transition, namely
from a low density system with broken chiral symmetry to a high density system where chiral 
symmetry is restored. 
The respective density is denoted by $\rho_{\rm chiral}$.
For densities larger than $\rho_{\rm chiral}$ the Goldstone bosons which characterize the 
chirally broken phase are no longer stable but can decay into quark-antiquark pairs.
If $\rho_{\rm chiral}$ was larger than $\rho_{\rm deconf}$ 
the following scenario would be conceivable: There would be three phases, namely 
(i) a hadronic, i.e.~confined phase at low densities, 
(ii) a phase where quarks are deconfined but massive (in this phase e.g.~pions
would still appear as bound states), 
and (iii) a high density phase where quarks are
deconfined and their masses are so low that all mesons can decay into quarks. 
Had we neglected all current quark masses, the quarks in the third phase would be 
exactly massless.
With finite current quark masses, however, the constituent quark masses keep on dropping
with rising density in the third phase (cf.~Fig.~\ref{fig:mRho}). 
This definitely interesting scenario with three phases is not realized in our model.
It turns out that the deconfinement phase transition happens far beyond the chiral transition,
i.e.~$\rho_{\rm chiral} < \rho_{\rm deconf}$. Therefore, only the phases (i) and (iii)
appear here. 

\begin{figure}
\centerline{\epsfig{file=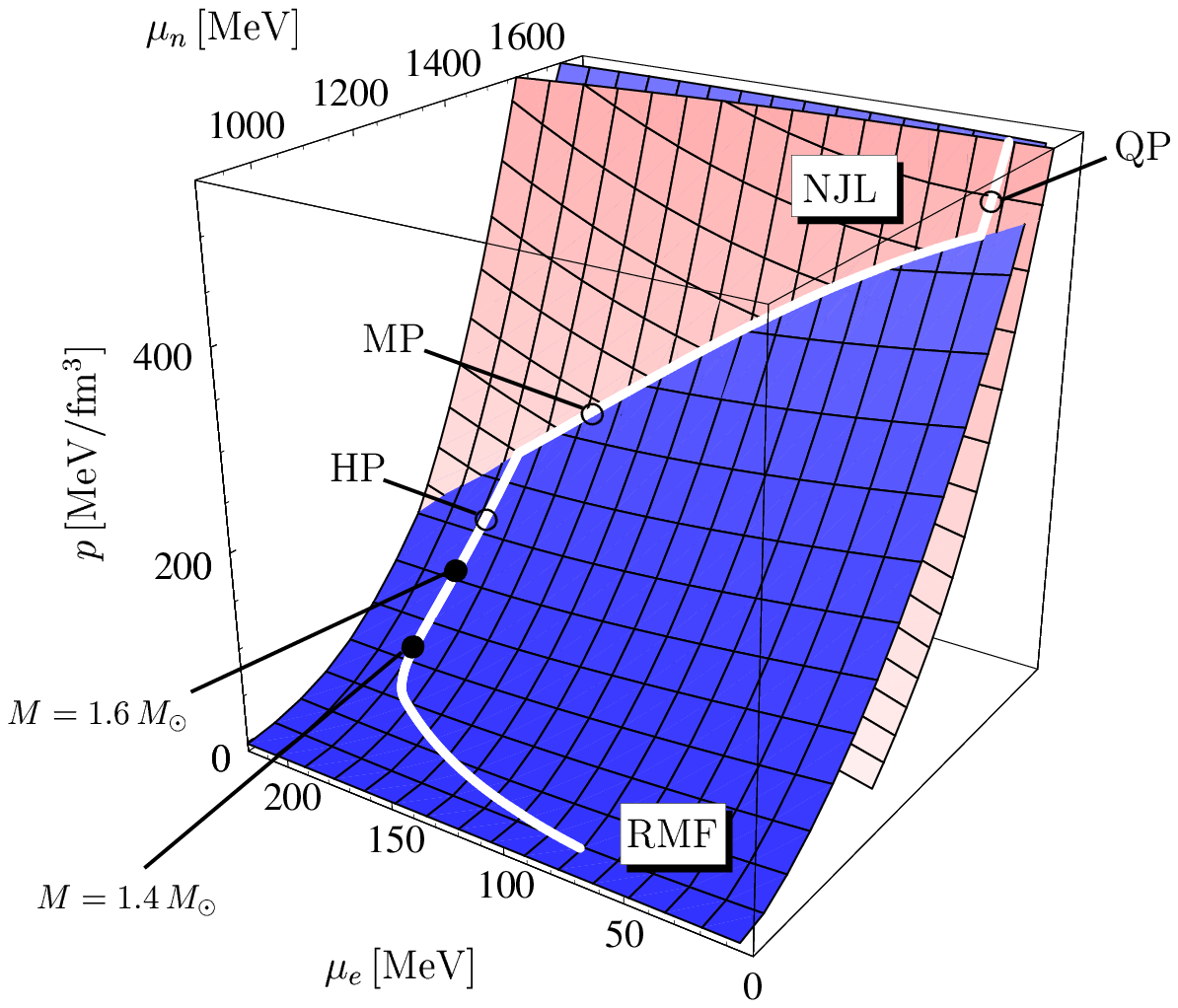,width=11cm}}
\caption{Gibbs phase construction of a two-component system. Plotted is the pressure surface 
of the hadronic
phase (RMF) and of the quark phase (NJL) as a function of the two independent
chemical potentials $\mu_n$, $\mu_e$. EOS of the HP is GPS, EOS of the QP uses the NJL model.
The white lines HP and QP show the
pressure of the hadronic and the quark phase under the condition of charge 
neutrality. At low pressure matter is in its charge neutral HP. 
The intersection curve MP corresponds to the mixed phase. This curve is
the solution of the Gibbs condition (\ref{GibbsCondition}). 
At very high pressure matter consists of a pure QP. Also shown are the central pressures of a
typical $M=1.4\,M_\odot$ neutron star and of an $M=1.6\,M_\odot$ star. 
($M_\odot$ denotes the mass of the sun.) Clearly neither achieves
a central pressure necessary to undergo a phase transition to deconfined matter.}
\label{p3D}
\end{figure}

In principle, since we assume the deconfinement phase transition to be of first 
order these two phases can coexist in a mixed phase. 
Indeed, it was first pointed out by Glendenning that beside a HP and a QP
also this mixed phase (MP) of quark and hadronic matter
may exist inside neutron stars \cite{GlenBook,Glen92}. (For a discussion of the geometrical structure
of the MP and its consequences for the properties of neutron stars see \cite{GlenBook}.)
This possibility was not realized in previous
calculations due to an inadequate treatment of neutron star matter as a one-component system
(one which can be parameterized by only one chemical potential). As we have already discussed,
the treatment of neutron star matter as a charge neutral phase in weak equilibrium indeed reduces
the number of independent chemical potentials to one. 
But the essential point is that - if a MP exists - charge neutrality can be achieved in
this phase e.g.~with a positively charged amount of hadronic matter and a negatively charged 
amount of quark matter.
Therefore it is not justified to require charge neutrality in both phases separately. 
In doing so
we would ``freeze out'' a degree of freedom which in principle could be exploited in the MP 
by 
rearranging electric charge between both phases to reach ``global'' charge neutrality. 
A correct treatment of the phase transition therefore
only requires both phases to be in weak equilibrium, i.e.~both phases still 
depend on two independent
chemical potentials. We have chosen the pair ($\mu_n$, $\mu_e$). Such a system is called 
a two-component system. 
The Gibbs condition for mechanical and chemical equilibrium at zero temperature between
both phases of the two-component system reads
\begin{equation}\label{GibbsCondition}
p_{HP}(\mu_n, \mu_e) = p_{QP}(\mu_n, \mu_e) = p_{MP}.
\end{equation}
Using Eq.~(\ref{GibbsCondition}) we can calculate the equilibrium chemical potentials of the MP where 
$p_{HP}=p_{QP}$ holds. Fig.\,\ref{p3D} illustrates this calculation. 
%
%
%
%
The HP$\rightarrow$MP phase transition takes place if the pressure of the charge neutral HP (white line)
meets the pressure surface of the QP (NJL). 
Up to this point
the pressure of the QP is below the pressure of the HP making the HP the physically realized one.
At higher pressure the physically realized phase follows the MP curve 
which is given by the 
Gibbs condition (\ref{GibbsCondition}). Finally the MP curve meets the charge neutral QP curve
(white line) and the pressure of the QP is above the pressure of the HP, 
making the QP the physically realized one.
For every point on the MP curve one now can calculate the volume proportion
\begin{equation} 
\chi = \frac{V_{QP}}{V_{QP}+V_{HP}}
\end{equation}
occupied by quark matter in the MP by imposing the condition of global charge neutrality of the MP
\begin{equation} \label{globalcharge}
\chi \, \rho_c^{QP} + (1-\chi) \, \rho_c^{HP} = 0.
\end{equation}
Here $\rho_c^{QP}$ and $\rho_c^{HP}$ denote the respective charge densities.
From this, the energy density $\epsilon$ of the MP can be calculated by
\begin{equation}\label{epsilonQP}
\epsilon_{MP} = \chi \, \epsilon_{QP} + (1-\chi) \, \epsilon_{HP}.
\end{equation}
Along the MP curve the volume proportion occupied by quark matter is monotonically 
increasing from $\chi=0$ 
to $\chi=1$ where the  transition to the pure QP takes place. 

\begin{figure}[ht]
\centerline{\epsfig{file=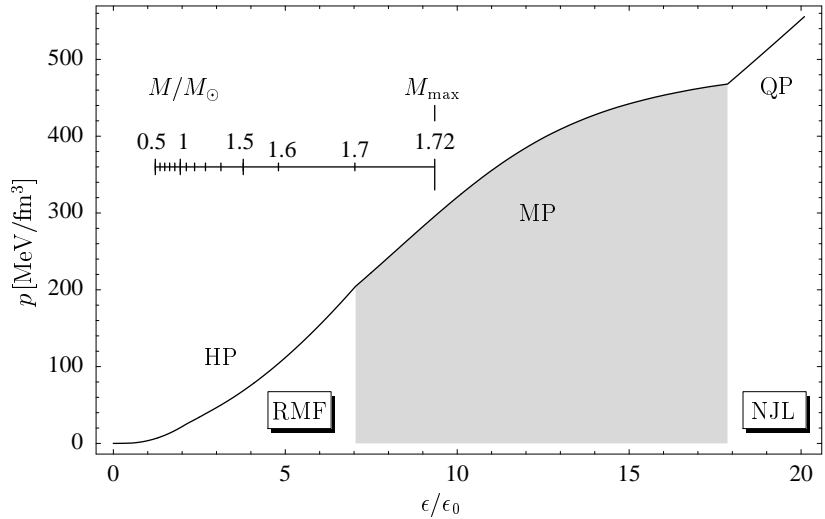,width=\figwidth}}
\caption{EOS in the form pressure versus energy density using GPS for the 
hadronic EOS. 
The shaded region (MP) corresponds to the mixed phase part of the EOS. 
\mbox{$\epsilon_0=140\,$MeV/fm$^3$}. 
The mass scale in the plot shows the respective neutron star mass if the corresponding energy density 
is regarded as the central density of the neutron star.
Obviously around the typical neutron star mass of $1.4\,M_\odot$ the neutron stars consist
solely of hadronic matter. $M_\odot$ denotes the mass of the sun.}
\label{pEpsGPS}
\end{figure}

Taking (i) the charge neutral EOS of the HP at low densities (Sec.~\ref{sec:HP}), (ii) 
Eq.~(\ref{GibbsCondition}), (\ref{globalcharge}), and (\ref{epsilonQP}) for the MP, 
and (iii) the charge neutral EOS of the QP (Sec.~\ref{sec:QP}) 
we can construct the full EOS in the form $p=p(\epsilon)$. 
For simplicity we denote this EOS as the {\em hybrid star} EOS.
Fig.~\ref{pEpsGPS} shows this EOS if we apply GPS for the HP EOS.
%
%
%
The hybrid star EOS consists of three distinct parts. At low densities 
($\epsilon\lapprox7\,\epsilon_0$) matter is 
still in its confined HP. At ($\epsilon\approx7\,\epsilon_0$) the first droplets of deconfined quark 
matter appear. Above this density matter is composed of a mixed phase of hadronic and quark matter.
This MP part of the EOS is shaded gray.
Only at unaccessible high densities ($\epsilon\gapprox18\,\epsilon_0$) matter 
consists of a pure QP. The preceding statements refer to the use of GPS for the EOS of the HP. 
Concerning all the other variants of RMF used here (TM1, TM2, GL85) we have found that the 
HP$\rightarrow$MP transition does not appear below 
$\epsilon \approx 10 \, \epsilon_0$. As we will discuss below
such high energy densities cannot be reached inside a stable neutron star which is described by
one of these EOS. 

At this point we should note the essential difference between the treatment of neutron star matter
as a one- and a two-component system (cf.~\cite{Glen92}). 
While the former one leads to the well known phase transition
with a constant pressure MP (like in the familiar liquid-gas phase transition of water), we can see in
Fig.\,\ref{pEpsGPS} that the pressure is monotonically increasing even in the MP if we apply
the correct two-component treatment. This has an important consequence on the structure
of the neutron star. 
Since we know from the equations of hydrostatic equilibrium -- the 
Tolman-Oppenheimer-Volkoff (TOV) equations \cite{OppeVolk39} -- that the pressure has to increase
if we go deeper into the star, a constant pressure MP is strictly excluded from the star while a 
MP with increasing pressure can (in principle) occupy a finite range inside the 
star. 

\begin{figure}[ht]
\centerline{\epsfig{file=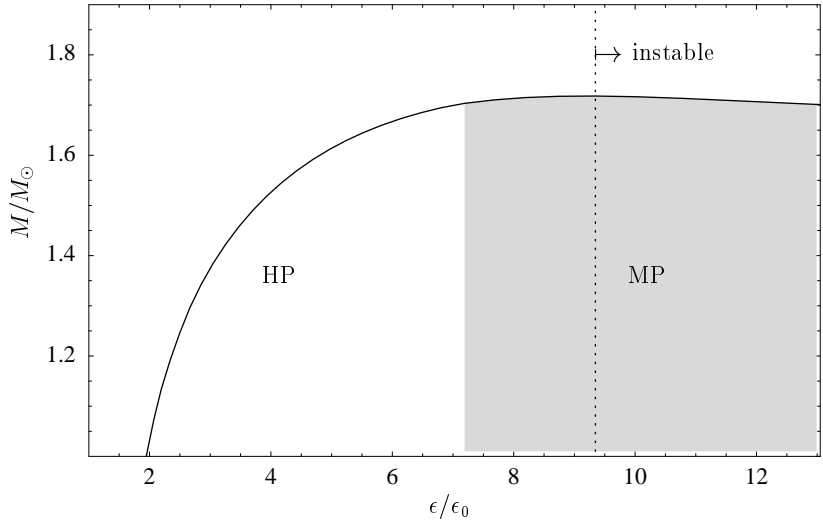,width=\figwidth}}
\caption{Mass of the neutron star as a function of the central energy density. Above a
central energy density of $\epsilon_{\rm crit} \approx 9 \, \epsilon_0$ where the maximum mass of the
neutron star is reached the system becomes instable with respect to radial modes of 
oscillations. The shaded region (MP) corresponds to stars which possess a mixed phase in 
their center. The hadronic part of the EOS uses GPS. 
\mbox{$\epsilon_0=140\,$MeV/fm$^3$}. 
$M_\odot$ denotes the mass of the sun.}
\label{fig:MEps}
\end{figure}

To see if the
densities inside a neutron star are high enough to establish a MP or a QP in its center we have to 
solve
the TOV equations with a specified hybrid star EOS following from our phase transition calculation.
From the solutions of the TOV equations we get a relation between the central energy density
(or central pressure) and the mass of the neutron star (cf.~Fig.~\ref{fig:MEps}).
The maximum possible central energy density 
(the critical energy density $\epsilon_{\rm crit}$) is reached
at the maximum mass that is supported by the Fermi pressure of the particular hybrid star EOS.
Above this critical density the neutron star gets instable with respect to radial modes of 
oscillations \cite{GlenBook}.
We have applied the four HP EOS's (denoted by GPS, TM1, TM2 and GL85) to calculate the four
corresponding hybrid star EOS's. (The one for GPS is shown in Fig.\,\ref{pEpsGPS}.) We found
that in {\em no} EOS the central energy density of a typical $M\approx1.4\,M_\odot$ neutron 
star is large
enough for a deconfinement phase transition. (Here $M_\odot$ denotes the mass of 
the sun.) The corresponding neutron stars are purely made of
hadronic matter (HP). In Fig.~\ref{pEpsGPS} where GPS is used also the neutron star masses are shown 
as a function of the
central energy density. There the central energy density of a $M=1.4\,M_\odot$ is about
$\epsilon\approx3\,\epsilon_0$ which is clearly below $\epsilon\approx7\,\epsilon_0$ which is
at least necessary to yield a MP core. (This is also shown in the context of the Gibbs construction 
in Fig.\,\ref{p3D} where the central pressure of a $M=1.4\,M_\odot$ and of a
$M=1.6\,M_\odot$ neutron star is marked.) In Fig.~\ref{pEpsGPS} we can see that 
only near the maximum mass of $M\approx1.72\,M_\odot$ neutron stars with a 
MP core are possible. This, however, 
only holds for the GPS hybrid star EOS and only in a quite narrow mass range from
$M\approx1.7\!-\!1.72\,M_\odot$. In the density range up to the critical density
all other EOS's (TM1, TM2, GL85) do not show a phase transition at all. The critical energy densities
for these EOS's are in the range of $\epsilon_{\rm crit} \approx 5 \!-\! 6 \, \epsilon_0$ 
while the densities for the HP$\rightarrow$MP transition are above 
$\epsilon \approx 10 \, \epsilon_0$. (The corresponding maximum masses are 
$M \approx 1.5 \! - \! 1.8 \, M_\odot$.)
From this we conclude that 
within the model constructed here the appearance
of deconfined quark matter in the center of neutron stars turns out to be very unlikely.

\section{Summary and conclusions}  \label{sec:sum}

We have studied the possible phase transition inside neutron stars from confined
to deconfined matter.  For the description of the quark phase we have utilized the NJL model
which respects chiral symmetry and yields dynamically generated quark masses via the 
effect of spontaneous chiral symmetry breaking.
We found that the appearance of deconfined quark matter in the
center of a neutron star appears to be very unlikely, for most of the studied hadronic
EOS's even impossible. The ultimate reason for that effect is the high value of the
effective strange quark mass which turns out to be much higher than its current mass value
in the whole relevant density range (cf.~Fig.~\ref{fig:mRho}). 
This finding, of course, is based on several assumptions
which need not necessarily be correct. In lack of an EOS based on a full QCD calculation
at zero temperature and finite nuclear density, we had to rely on simpler models for the
EOS in different density regimes. 

Concerning the low density regime we have used various
relativistic mean field (RMF) models. To some degree the use of different variants
of the RMF model should reflect the uncertainties of this approach. 
These models are generalizations of the Walecka
model \cite{SW88} which describes the hadronic ground state of nuclear matter 
at density $\rho_0$ quite successfully. At somewhat higher densities the used RMF models
deal with hyperons as additional degrees of freedom. Since these RMF models do not have
any explicit quark degrees of freedom we expect them to become unreliable at high
densities where the confinement forces are screened and the hadrons dissolve into quarks.

To describe this high density regime we have utilized the Nambu--Jona-Lasinio (NJL) model
in its three-flavor extension. The merits of the NJL model are (at least) twofold:
For the vacuum case, it gives a reasonable description of spontaneous chiral symmetry 
breaking and of the spectrum of the low lying mesons. For sufficiently high density and/or 
temperature, the NJL model exhibits the restoration of chiral symmetry. 
A shortcoming of the NJL model is that it does not confine quarks, i.e.~there
is no mechanism which prevents the propagation of a single quark in vacuum. Therefore
in an NJL model calculation the quarks significantly contribute to the EOS also at low 
densities.\footnote{Actually in the mean field approximation the quarks are the only 
degrees of freedom which contribute. For a generalization to include meson states as 
RPA modes in the NJL EOS see \cite{HKZV94,ZHK94}.} This was the 
ultimate reason why we
considered the NJL model only in the high density regime where confinement is supposed to
be absent anyway while utilizing the hadronic RMF models to describe the confined phase. 
On the other hand, we should recall that energy density and pressure of the NJL model
were determined such that both vanish at zero density, i.e.~in a regime where we have not
utilized the NJL model afterwards. This procedure determines the effective bag pressure 
$B_{\rm eff}$ given in (\ref{eq:defeffbag}) by fixing $B_0$ (\ref{eq:defb0}) to 
$(217.6 \,\mbox{MeV})^{4}$. Clearly, this procedure is somewhat unsatisfying since
the effective bag pressure $B_{\rm eff}$ influences
the EOS and therefore the onset of the phase transition. 
Indeed, if we reduce $B_0$ by only $5 \! - \! 10$\% 
by hand from its original value of $(217.6 \,\mbox{MeV})^{4}$ 
we already observe drastic changes in the phase structure of the neutron star favoring 
deconfined quark matter. 
On the other hand, the physical requirement that any model
should yield vanishing energy density and pressure in vacuum is the only way to uniquely
determine the EOS of the NJL model without any further assumptions.

Possible alternatives to the use of the NJL model for the description of the deconfined
quark matter are the MIT bag model \cite{GlenBook} and the extended effective mass
bag model \cite{Sche97,Sche98}. The latter includes medium effects due to one-gluon
exchange which rise with density (while the effective masses of the NJL model decrease). 
As already discussed at the end of Sec.~\ref{sec:QP} for the regime of very high densities
such a resummed perturbation theory might be more adequate. 
Furthermore, MIT bag models can be very useful in interpreting more involved models
like the NJL model in terms of simple physical quantities like the bag constant and 
the quark masses. Therefore we have frequently compared our NJL model results with the
MIT bag model in Sec.~\ref{sec:QP}. 
For a further discussion of the MIT bag model and its application to neutron stars
see \cite{Sche99}.

The distinct feature of the NJL model is that 
nonperturbative effects are still present beyond the phase transition point. 
It is reasonable to consider such effects since it was found in lattice calculations 
\cite{Dehwa2,LaerQM} that for QCD at finite
temperature the EOS beyond the phase transition point can neither be properly described by 
a free gas of quarks and gluons nor by QCD perturbation theory \cite{ZK95,BN96}. 
Presumably this holds also for the finite density regime. On the mean field level the
most prominent nonperturbative feature of the NJL model which remains present beyond the
phase transition point is the constituent strange quark mass which is much larger than 
the current strange quark mass in the whole relevant density regime 
(cf.~Fig.~\ref{fig:mRho}). 
This high strange quark mass has turned out to be crucial for the phase transition.
As we have shown above the EOS of the QP can be reasonably well
approximated by an MIT bag EOS up to 5 times $\rho_0$ using a comparatively low bag
constant of $B_{\rm MIT}^{1/4} = 170\,$MeV and an effective strange quark mass of
$m_s^{\rm MIT} = 465 \,$MeV. These are the plateau values of the corresponding quantities
in the NJL model calculations shown in Figs.~\ref{fig:mRho} and \ref{fig:BRho}. 
In Figs.~\ref{fig:EARho}-\ref{fig:pEps} the curves labeled by MIT use these values for
the bag constant and the effective strange quark mass.
Using such a bag constant in connection with the {\it current} strange quark 
mass would allow the existence of a QP inside a neutron star
\cite{GlenBook,Sche99}. This, however, does not remain true once a much higher effective
strange quark mass is used. 
Qualitatively, the chain of arguments is that a higher mass leads to a lower pressure 
(cf.~Fig.~\ref{fig:pMub}). This disfavors the quark phase in the Gibbs 
construction, i.e.~shifts the phase transition point to higher densities. 
This is the reason why in our calculations the existence of quark matter in the 
center of a neutron star is (nearly)
excluded. Especially for typical neutron stars with masses 
$M \approx 1.4 M_\odot$ the central energy density is far below the deconfinement
phase transition density (cf.~Fig.~\ref{fig:pEps}). 
This finding is independent of the choice of the version of the 
RMF model. This suggests that it is the NJL model with its large strange quark mass
which defers the onset of the deconfinement phase transition rather than the 
modeling of the hadronic phase. Throughout this work we have used the NJL parameter
set of \cite{RKH96}. We have also explored the set given in \cite{Ku89} with
$\Lambda = 631\,$MeV, $G \Lambda^2 = 1.830$, $K \Lambda^5 = 9.19$, $m_q=5.5 \,$MeV,
and $m_s = 138 \,$MeV. The results are very similar to the ones presented here. 

For simplicity we have treated in the present work the NJL model in the Hartree 
approximation. In principle, going beyond
the mean field approximation might influence the order of the chiral phase 
transition
(for related work towards that direction for the two-flavor case cf.~\cite{BJW98}). 
If it turned out that this would result in a strong first order phase transition 
then the effective
strange quark mass might change more drastically and in the region of interest would be 
perhaps much lower than in the case studied in the present work. This would favor the
appearance of quark matter in the interior of neutron stars. Clearly, it would be interesting
to study how a more involved treatment of the NJL model beyond the Hartree approximation
would influence our findings presented here. This, however, is beyond the scope of the
present work.

\medskip

{\bf Acknowledgments:} 
The authors thank C.~Greiner and M.H.~Thoma for helpful discussions 
and for reading the manuscript. We also acknowledge discussions with M.~Hanauske.


\begin{thebibliography}{99}

\bibitem{ShapiroBook}S.L.~Shapiro and S.A.~Teukolsky, Black Holes, 
White Dwarfs, and Neutron Stars (John Wiley \& Sons, N.Y., 1983).

\bibitem{GlenBook}N.K.~Glendenning, Compact Stars (Springer-Verlag, 1997).

\bibitem{MITbag}A.~Chodos, R.L.~Jaffe, K.~Johnson, C.B.~Thorn, and V.F.~Weisskopf, 
Phys. Rev.~D9 (1974) 3471; \\
A.~Chodos, R.L.~Jaffe, K.~Johnson, and C.B.~Thorn, Phys.~Rev.~D10 (1974) 2599.
 
\bibitem{FarhiJaffe84}E.~Farhi and R.L.~Jaffe, Phys.~Rev.~D30 (1984) 2379.

\bibitem{Bodm71}A.R.~Bodmer, Phys.~Rev.~D4 (1971) 1601.

\bibitem{Witte84}E.~Witten, Phys.~Rev.~D30 (1984) 272.

\bibitem{MadsHaen91}Strange Quark Matter in Physics and Astrophysics, edited
by J.~Madsen and P.~Haensel, Nucl.~Phys.~B (Proc. Suppl.) 24B (1991).

\bibitem{BO98} M.~Buballa and M.~Oertel, hep-ph/9810529.

\bibitem{VW91} U.~Vogl and W.~Weise, Progr.~Part.~Nucl.~Phys.~27
                 (1991) 195.

\bibitem{Kl92} S.P.~Klevansky, Rev.~Mod.~Phys.~64 (1992) 649.

\bibitem{HK94} T.~Hatsuda and T.~Kunihiro, Phys.~Rep.~247 (1994) 221.

\bibitem{FTS95} R.J.~Furnstahl, H.-B.~Tang, and S.D.~Serot, Phys.~Rev.~C52 (1995) 1368; \\
R.J.~Furnstahl, S.D.~Serot, and H.-B.~Tang, Nucl.~Phys.~A615 (1997) 441; 
Erratum-ibid.~A640 (1998) 505.

\bibitem{PZ98} P.~Papazoglou, S.~Schramm, J.~Schaffner-Bielich, H.~St\"ocker, 
and W.~Greiner, Phys.~Rev.~C57 (1998) 2576; \\
P.~Papazoglou, D.~Zschiesche, S.~Schramm, J.~Schaffner-Bielich, H.~St\"ocker, 
and W.~Greiner, Phys.~Rev.~C59 (1999) 411.

\bibitem{Mu98} H.~M\"uller, nucl-th/9810055.

\bibitem{priv} M.~Hanauske, privat communication.

\bibitem{BPS}G.~Baym, C.J.~Pethick, and P.~Sutherland, Astrophys.~J.~170 (1971) 299; \\
R.P.~Feynman, N.~Metropolis, and E.~Teller, Phys.~Rev.~75 (1949) 1561;\\
G.~Baym, H.A.~Bethe, and C.J.~Pethick, Nucl.~Phys.~A175 (1971) 225.

\bibitem{STOS98} H.~Shen, H.~Toki, K.~Oyamatsu, and K.~Sumiyoshi, Nucl.~Phys.~A637 (1998) 435.

\bibitem{RMF}N.K.~Glendenning, F.~Weber, and S.A.~Moszkowski, 
Nucl.~Phys.~A572 (1994) 693; \\
J.I.~Kapusta and K.A.~Olive, Phys.~Rev.~Lett.~64 (1990) 13; \\ 
J.~Ellis, J.I.~Kapusta, and K.A.~Olive, Nucl.~Phys.~B348 (1991) 345.

\bibitem{Glen8287}N.K.~Glendenning, Phys.~Lett.~B114 (1982) 392; \\
N.K.~Glendenning, Z.~Phys.~A327 (1987) 295.

\bibitem{SchaMish96}J.~Schaffner and I.N.~Mishustin, Phys.~Rev.~C53 (1996) 1416.

\bibitem{Gosh95}S.K.~Gosh, S.C.~Phatak, and P.K.~Sahu, Z.~Phys.~A352 (1995) 457.

\bibitem{NJL} Y.~Nambu and G.~Jona-Lasinio, Phys.~Rev.~122 (1961) 345; 
Phys.~Rev.~124 (1961) 246.

\bibitem{RKH96} P.~Rehberg, S.P.~Klevansky, and J.~H\"ufner,
                  Phys.~Rev.~C53 (1996) 410.

\bibitem{HKZV94} J.~H\"ufner, S.P.~Klevansy, P.~Zhuang, and H.~Voss, 
Ann.~Phys.~234 (1994) 225.

\bibitem{ZHK94} P.~Zhuang, J.~H\"ufner, and S.P.~Klevansy, Nucl.~Phys.~A576 (1994) 525.

\bibitem{Sche97}K.~Schertler, C.~Greiner, and M.H.~Thoma, Nucl.~Phys.~A616 (1997) 659.

\bibitem{Sche98}K.~Schertler, C.~Greiner, P.K.~Sahu, and M.H.~Thoma, 
Nucl.~Phys.~A637 (1998) 451.

\bibitem{OppeVolk39}J.R. Oppenheimer and G.M. Volkoff, Phys. Rev. 55 (1939) 347.

\bibitem{Glen92} N.K.~Glendenning, Phys.~Rev.~D46 (1992) 1274.

\bibitem{SW88} B.D.~Serot and J.D.~Walecka, Adv.~Nucl.~Phys.~16 (1986) 1.

\bibitem{Sche99} K.~Schertler, J.~Schaffner-Bielich, C.~Greiner, and M.H.~Thoma, in 
preparation.

\bibitem{Dehwa2} C.~DeTar, in {\sl Quark-Gluon Plasma 2}, p.~1, edited by R.C.~Hwa 
(World Scientific, Singapore, 1995). 

\bibitem{LaerQM} E.~Laermann, Nucl.~Phys.~{A610} (1996) 1c.

\bibitem{ZK95} C.~Zhai and B.~Kastening, Phys.~Rev.~{D52} (1995) 7232.

\bibitem{BN96} E.~Braaten and A.~Nieto, Phys.~Rev.~{D53} (1996) 3421.

\bibitem{Ku89} T.~Kunihiro, Phys.~Lett.~B219 (1989) 363.

\bibitem{BJW98} J.~Berges, D.-U.~Jungnickel, and C.~Wetterich, hep-ph/9811347.

\end{thebibliography}
\end{document}